# *Electronic Energy Singularities of Weakly H-bonded Ammonium Dimer*

<u>Author</u>

Md Rejwan Ali, PhD
Dept of Material Science & Chemical Engineering
Stony Brook, New York
USA

**Corresponding author: Md Rejwan Ali*

*<alimd.rejwan@stonybrook.edu>*

## *Abstract*

Quantum and molecular mechanics based electronic energy studies of weak H-bonded ammonium dimer show distinctive feature in energy profile when computed by different QM methods contrast to MM methods. MM based MMFF and SYBYL methods show smoothly varying dihedral energy profile for torsion angle variation around weak N1-H5 held by H-bond strength of around 13 KJ/mol. All the QM based methods HF, B3LYP and MP2 show noisy and unstable torsion dependent electronic energy profile for H-bonded ammonium dimer. Exploring energy surface beyond bond length shows singularities and discontinuities. QM-based computation of dipole moment shows several discreet values with jumps and discontinuities with torsion angle variation for ammonium dimer. Also repeated computations and reverse torsion energy profile show persistent singularity feature observed in all standard QM techniques.

## *Introduction*

Over the years nature of the hydrogen bond in ammonium dimer has been well studied by both contested theoretical and experimental techniques. From structural and spectroscopic perspectives, ammonium dimer has been investigated with interesting findings and results [1,2,3]. Several different ammonium dimers work presented views with contrasting experimental and theoretical results [4,5,6]. However, recent studies reconfirm ammonium dimer to have weak H-bonding between N1 and H5 atoms (Fig.1) [7,8].

In this short report, before exploring the nature of the electronic energy profile for weak H-bond, we have computed bond length and energy for optimized ammonium dimer geometry within reasonable reproducible values with the computational tools used in the work and data shown in Table 1. As per recent reported works, such weak and unstable H-bonded dimers' pseudo-stable conformers, mostly generated via molecular mechanics method with a continuum of energy profile, are highly unlikely to have geometrical stability with torsion and must carry quantum break-point signature in electronic energy computation due to the intrinsic weak H-bond [9,10]. In such predicted geometrical instability with torsional rotation, H-bond breaking criticality should appear in ab initio molecular electronic energy computation rather than a type of periodic smooth energy function with dihedral angle dependence in MM-based calculations. The key governing factor in the current studied system is, of course, the weak H-bond in ammonium dimer and the fact that all structures from MM prediction are not viable by QM techniques due primarily to the onset of breaking condition of the dimer structure. As quantum mechanical bond-breaking conditions do not require connected MM-based molecular topology, and such energy results are inherent in the system and not due to any numerical convergence issues in electronic energy computation [11]. One interesting computational probe has been to explore the nature of electronic energy beyond the dimer bond length with full bond rotation in torsion space. The distinct feature of anisotropic energy criticalities computed by HF/6-31G+ compared to smooth MMFF energy surface is also indicative of geometrical criticality of weak H-bond beyond ammonium dimer bond length in as studied in range 2.3 to 2.8 Å.

## ***Computational Methodologies***

Before each torsion-based energy profiling task, the optimized geometry of the ammonium dimer has been obtained for both the QM and MM methods. For MM methods, SYBYL and MMFF94 have been applied, while ab initio methods have been done via standard HF, DFT and MP2 at theory levels 6-311G* and aug-cc-PVDz as implemented in SPARTAN 20 platform (URL: https://www.wavefun.com; Wavefunction, INC., Irvine, CA). Geometry optimization, bond length, bond angle, and dipole moment computed by SPARTAN tools have reasonably reproduced previous reported values [3,12,13,14,15].

The theoretical binding energy of NH3 dimer formation was estimated from the hydrogen bond well depth between N1 and H5 (Fig.1) atoms to ~ 13.0 KJ/mol without zero point energy correction [16,17]. A bond scan was performed with N1-H5 coordinate at 0.1 Å resolution between 1.5 to 8 Å. The dissociation energy has been estimated from the well depth using global minima and asymptotic unbounded energy threshold computed from the energy plot for each ab initio technique [18]. The results have been tabulated in the Table. 1.

The electronic energy singularity feature of ammonium dimer has been explored by HF/6-31G+ to compare with MMFF-based results. We have explored critical ranges both by MM and QM beyond the estimated bond length limit as in Table 1. With initial optimized geometry from each method, constrain distance between N1-H5 atoms was constrained, and the structure was again energy minimized. After the minimization step, the distance between N1-H5 atoms was reconfirmed before energy profile computation for N1-H5 bond full rotation at the same theory level as applied for initial geometry optimization. With both MM and QM tools, the energy surface has been explored in the range of 2.4 to 2.8 Å at 0.1 Å resolution. Dihedral scan has been done with 1° resolution. As seen in Table 1 and Figs. 6A and 6B, energy estimates from MM techniques are off from QM tools by a substantial difference. Surface energy plots were generated by converting cylindrical symmetry data (bond length r, torsion angle f and energy AU) to Cartesian format as per the GNUplot 3D graphing routine (URL link: http://www.gnuplot.info/). The energy surface is plotted in Figs. 6A and 6B.

## ***Results and Discussion***

### ***Table 1***

| Method | Dimer Binding Energy (KJ/mol) | N1H5 Bond Length (Å) | N6H5N1 Angle (Degree) | Dipole Moment (Debye) |
|---|---|---|---|---|
| HF/6-311G+ | 13.0 | 2.4 | 169.6 | 2.29 |
| B3LYP/6-311G+ | 17.6 | 2.2 | 176.9 | 2.45 |
| MP2/aug-cc-PVDz | 14.6 | 2.3 | 167.7 | 2.03 |
| MMFF | 42.7 | 1.1 | 109.8 | 3.53 |
| SYBYL | 36.4 | 1.6 | 101.1 | 3.76 |

All ab initio methods conclusively show that the molecular electronic energy carries a singularity feature in the ammonium dimer due to its weak H-bonding under torsion angle variation. While given molecular mechanics, all torsion-dependent conformers are geometrically feasible with energy continuum as observed in Figs 4A & B. However, such a continuum from MM-based results differs in quantum-based computations showing discontinuous molecular electronic energy and geometry computed by all standard methods, i.e., HF, B3YLP and MP2. Only in the limiting case QM energy profile should reduce to the MM energy, and MM always predicts bond stability around the torsion space if no clashes in atoms are present. However, energy surface beyond the bond limit in QM techniques does show singularities in energies due to weak H-bonding. Torsion-dependent dipole moment studies in Fig. 5 show several discrete dipole moment levels, and the pattern not surprising given the molecular geometry must reach criticality due to a weak H-bond. Interestingly, MM MM-based results show two distinct levels for dipole moment. While the results are interesting even from MM's computational perspective, it only highlight that certain discrete allowed geometries, including bond-breaking criticality, are stair-type consistent with its inherent quantum nature. In a broader view, these computational results can potentially be extended to other weakly H-bonded and Van der Waals (VDW) dimers with an expected anisotropic singularity in the energy surface beyond the weak H-bond length limit. From several previous studies, accurate QM-based energy surface and profile do show the distinctive energy feature in contrast to MM-based smooth energy surface or well-behaved dihedral energy function for weakly bonded dimers.

## ***Conclusion***

For internal rotation, quantized electronic energy comes as a natural solution when a torsion-like potential is plugged into the Schrodinger time-independent equation, reducing it to the Mathieu equation in computing energy. In contrast to results in molecular mechanics, quantized energy is what nature has over the energy continuum. Extensive studies on different systems conclusively show that the electronic energy singularities are an inherent quantum mechanical feature that suggests the relevance of current ab initio methods in exploring critical molecular geometry beyond bond length that have been consistently observed in weak H-bonded and also in VDW dimers. Singly H-bonded weak dimers with dissociation energy in the range of 10 to 25 KJ/mol can readily show such electronic energy feature in all standard ab initio computations.

Finally, the continuum of geometry is not allowed under quantum mechanics under torsion, while force field or molecular mechanics do not have such restrictions as long as there are no steric clashes among the atoms. Energy singularities in weak H-bond or VDW dimers and associated geometrical criticality are purely due to the quantum nature of bond break-up, as these results are distinct from molecular mechanics or force-field-based calculations.

## Acknowledgements

The author acknowledges many stimulating scientific discussions with Dr. Mihaly Mezei of the Icahn School of Medicine at Mount Sinai. Computational support from NSF-supported Texas Advanced Computing Center (TACC) resources is to be mentioned. Finally author is thankful to Mr. Sean Ohlinger of Wavefunction for license continuation with SPARTAN 20 software.

## *Data Availability*

The details of the hypothesis and the dataset that support the findings of this study are available in the Zenodo repository in the following link: https://zenodo.org/records/12730902.

**REFRENCES**

# *Electronic Energy Singularities of Weakly H-bonded Ammonium Dimer*

## Figure Slides

**Fig.1A & 1B: Optimized Geometry of Ammonium Dimer Computed by HF/6-311G*(L) and MP2/cc-PVDz (R)**

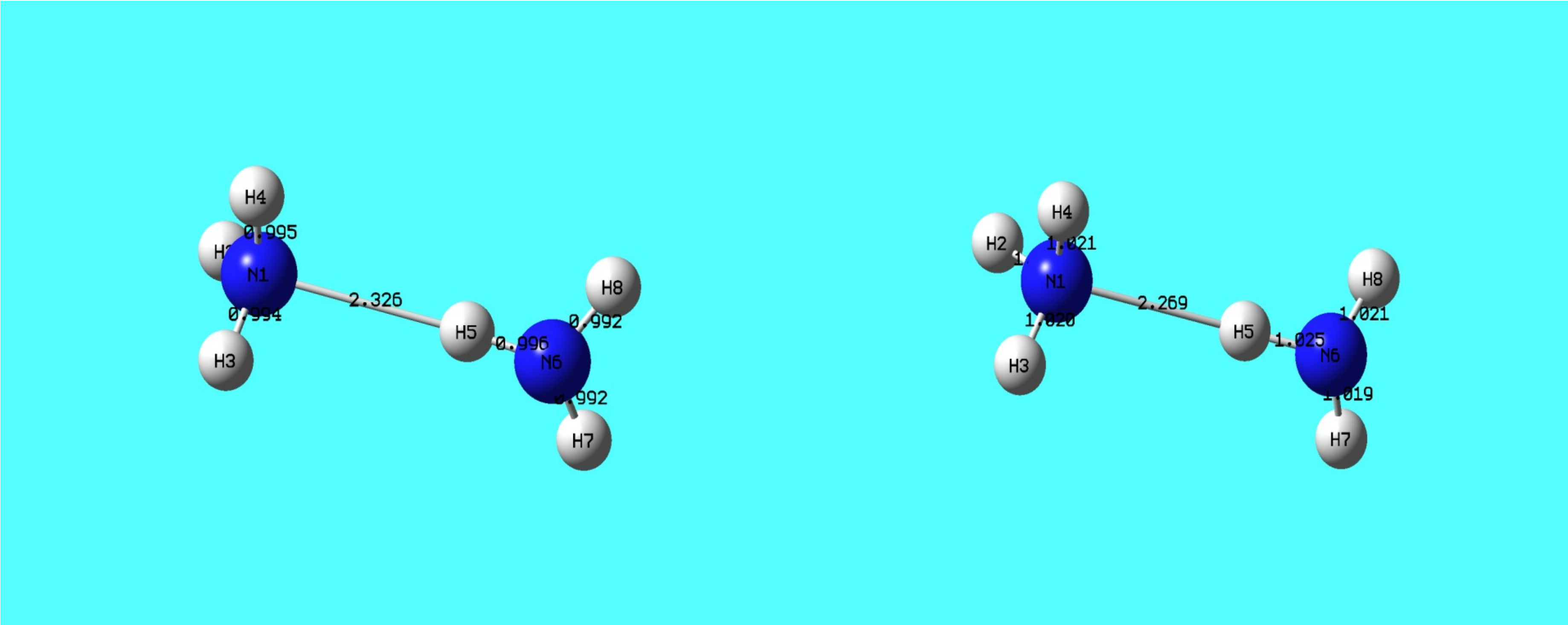

## Fig.2A & 2B: Ammonium Dimer Electronic Energy Profile Showing Noisy Feature with Several Discontinuities. Computed by HF/6-311G*(L) and MP2/aug-cc-PVDz (R)

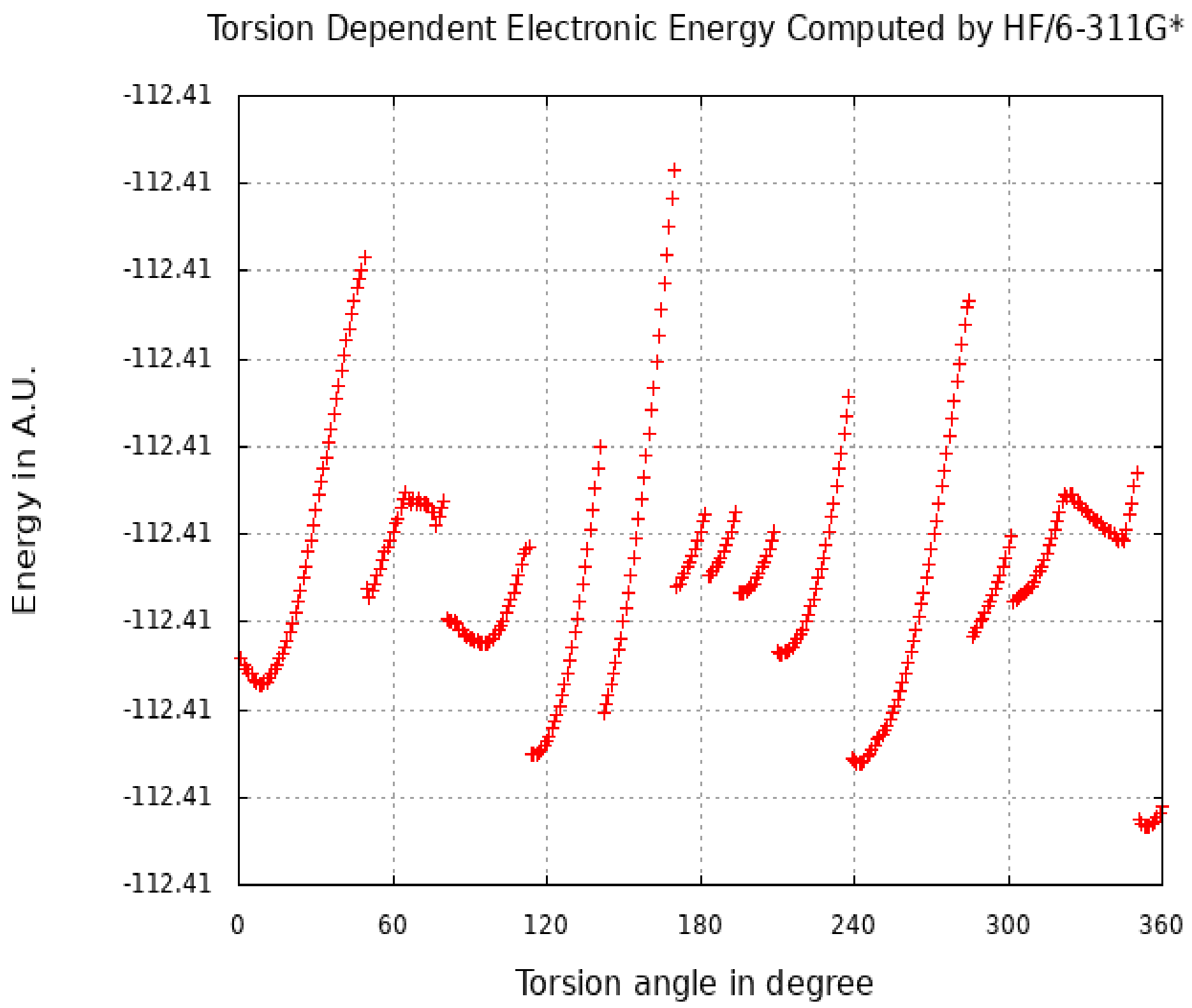

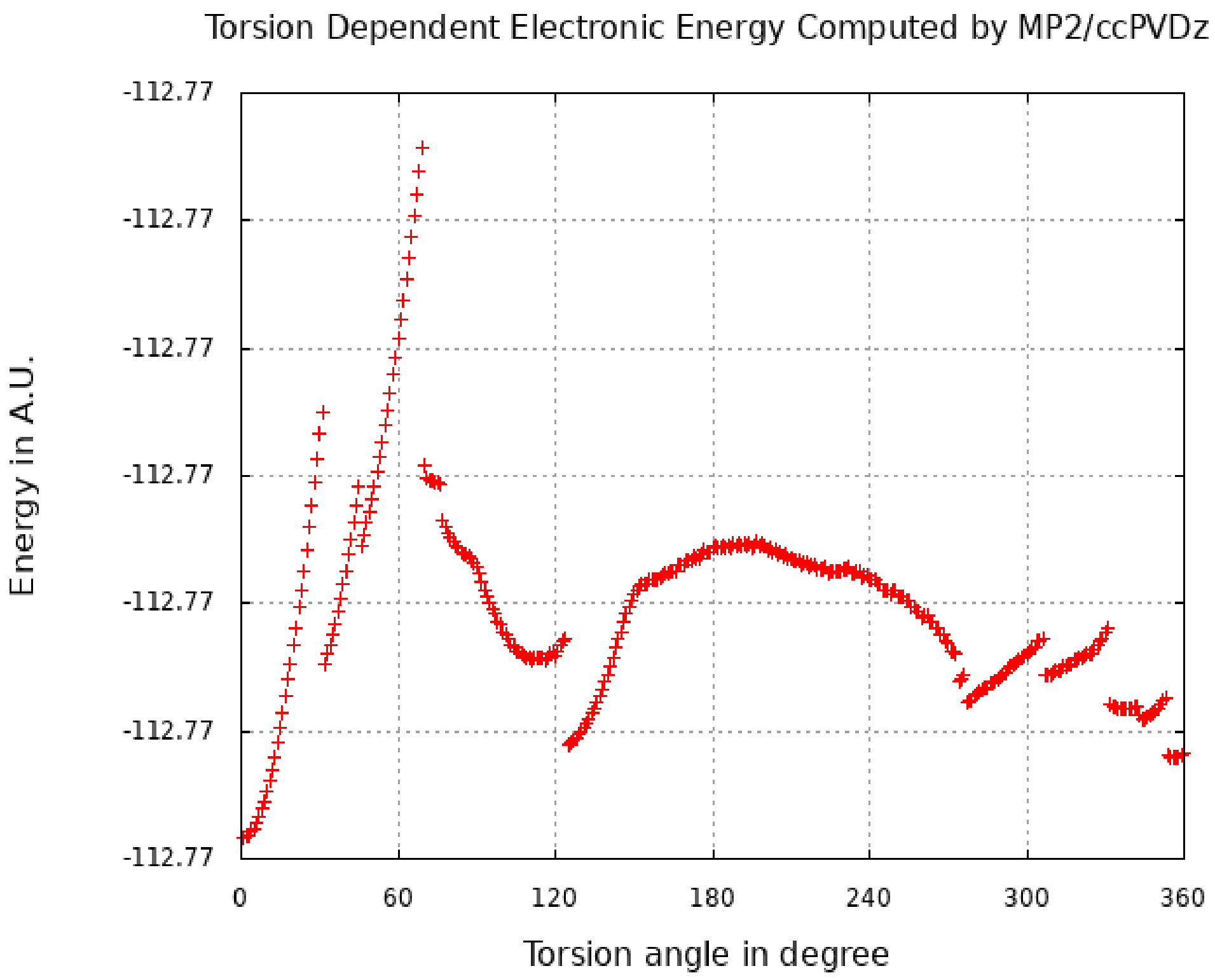

# Fig.3A & 3B: Ammonium Dimer Electronic Energy Profile Showing Computed Forward (L) and Backwards (R) by B3LYP/6-311G*(L) Basis

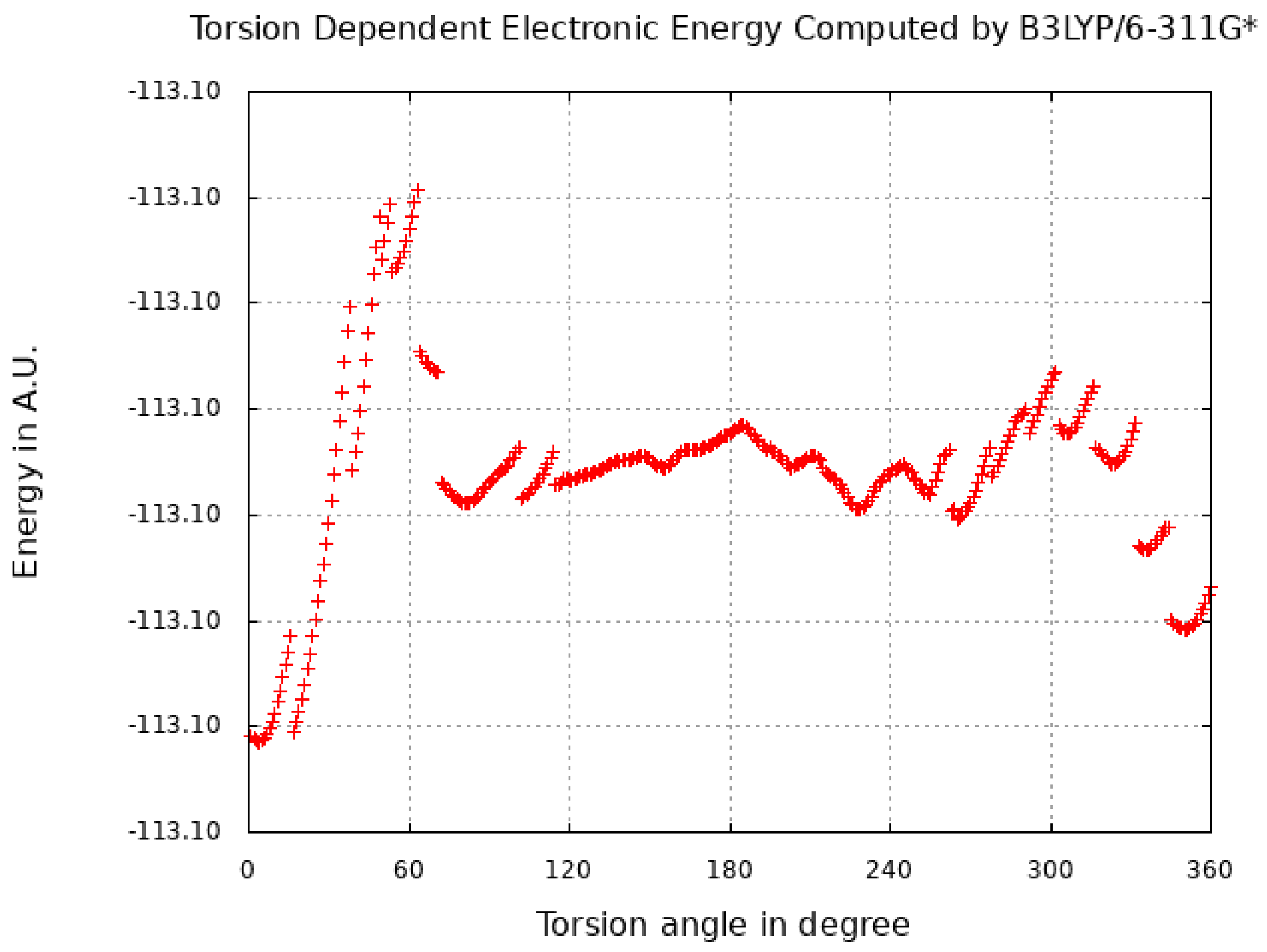

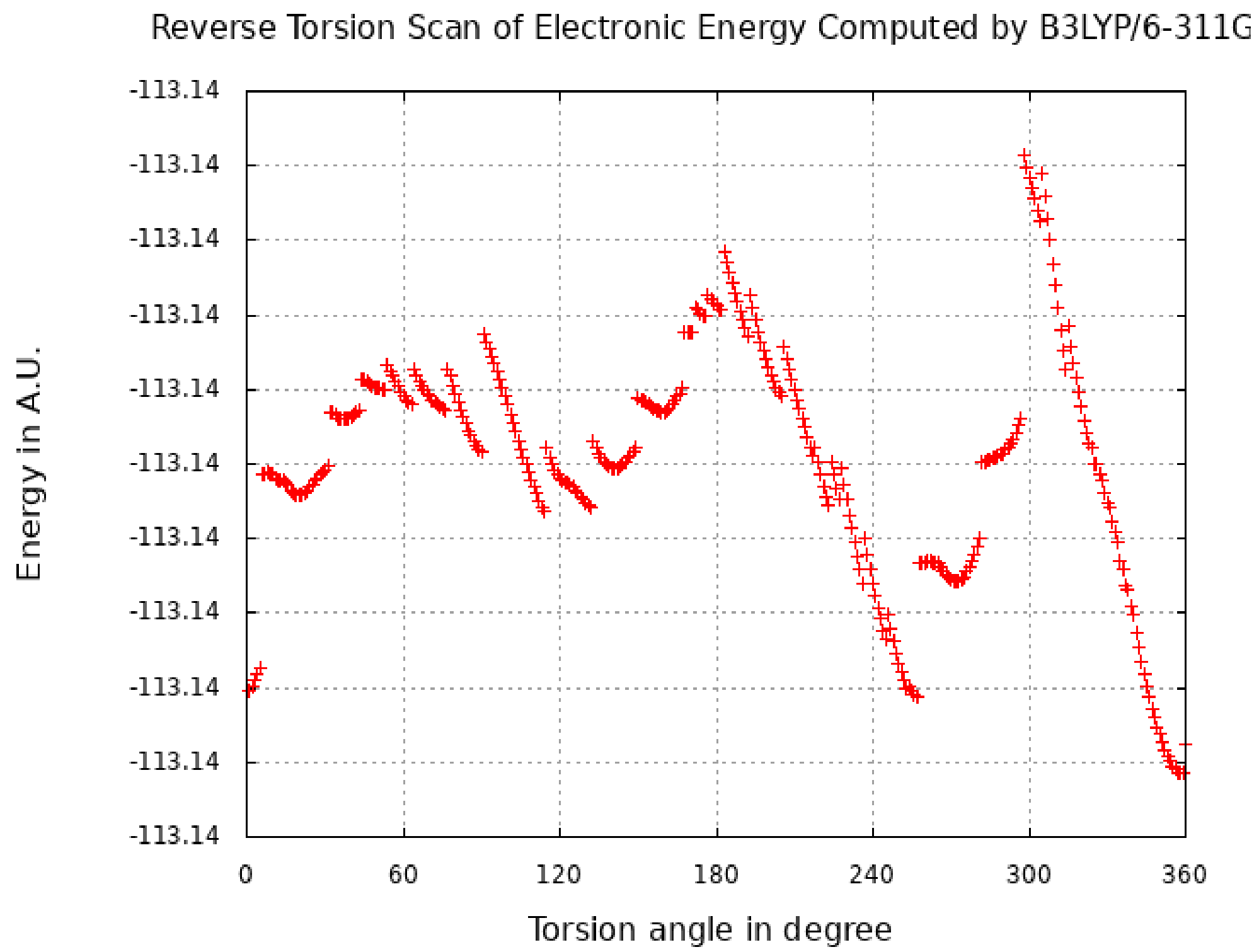

# Fig.4A & 4B: Torsion Energy Profile Around Weak H-bond Computed by MMFF(L) and SYBYL (R) Methods.

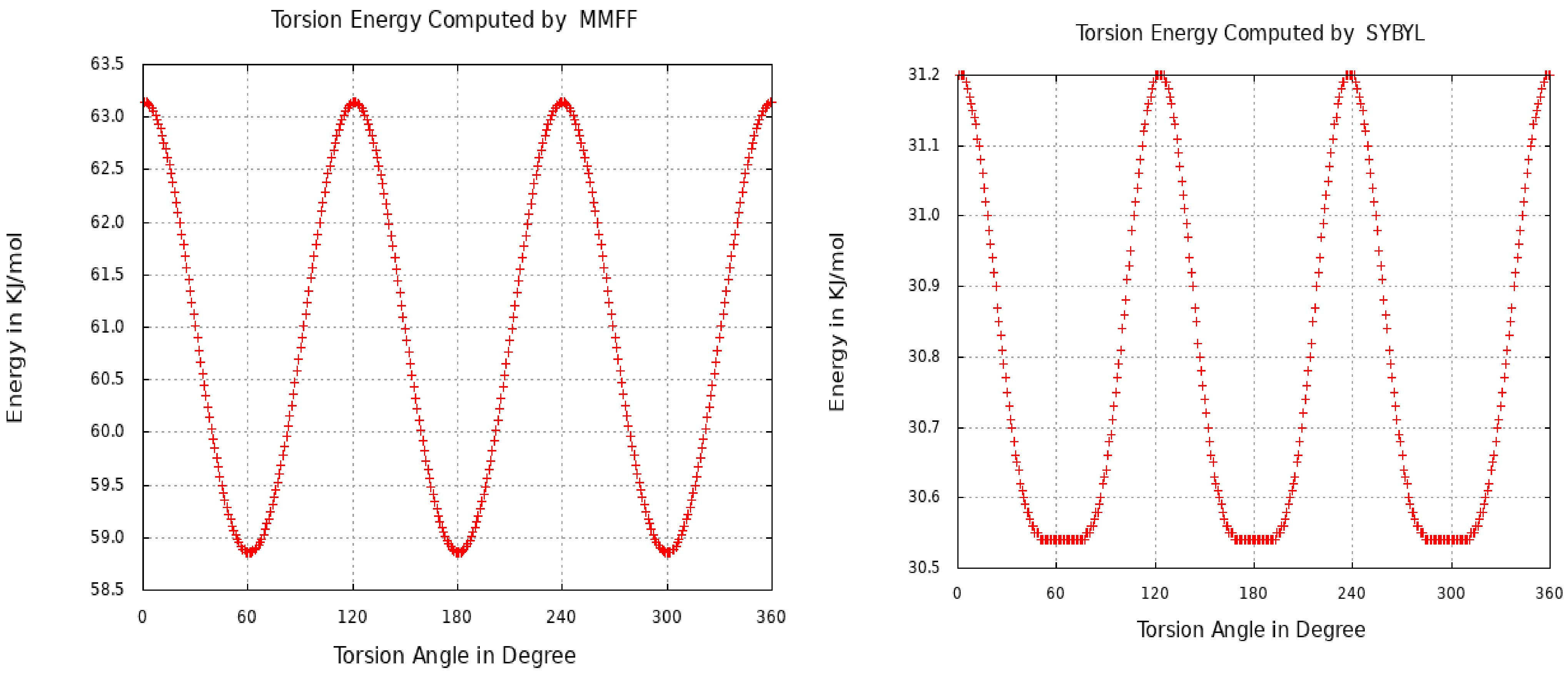

**Fig.5A, 5B & 5C: Dipole Moment Computed by MM(L) and QM Techniques (M & R). QM Computations Show Many Dipole Levels compared to Distinct Two Levels in MM Method.**

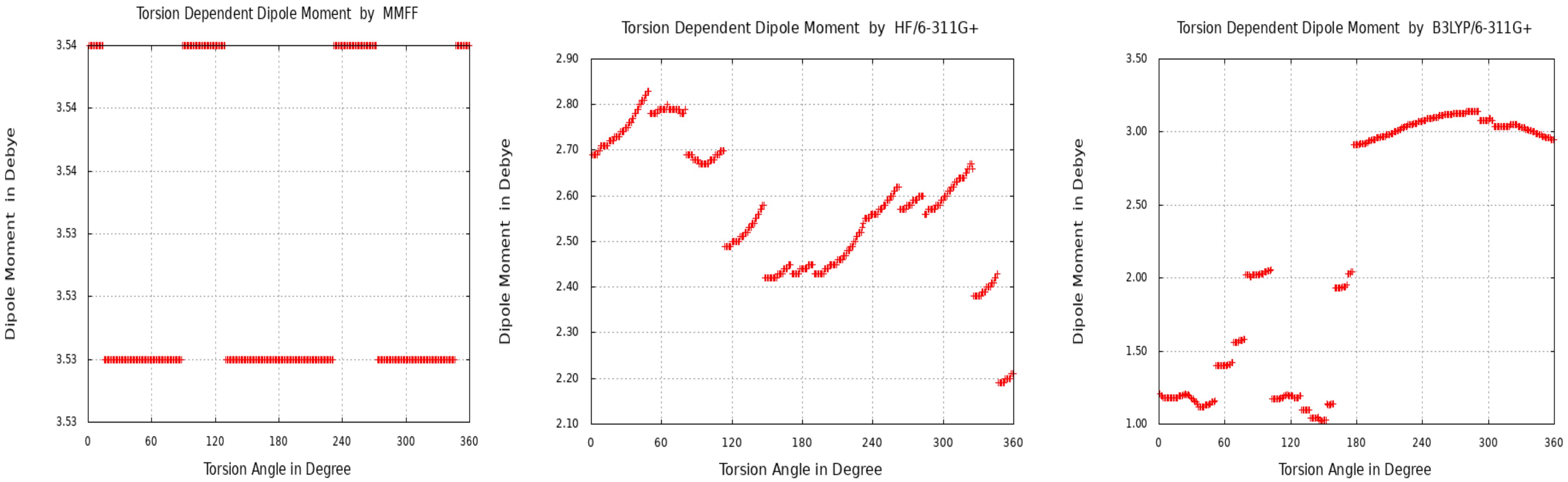

# Fig.6A, 6B: ANISOTROPIC ENERGY SINGULARITIES OF AMMONIUM DIMER DUE TO WEAK H-BOND TORSION BY HF/6-31G+ (L) COMPARED SMOOTH ISOTROPIC ENERGY SURFACE BY MMFF (R).

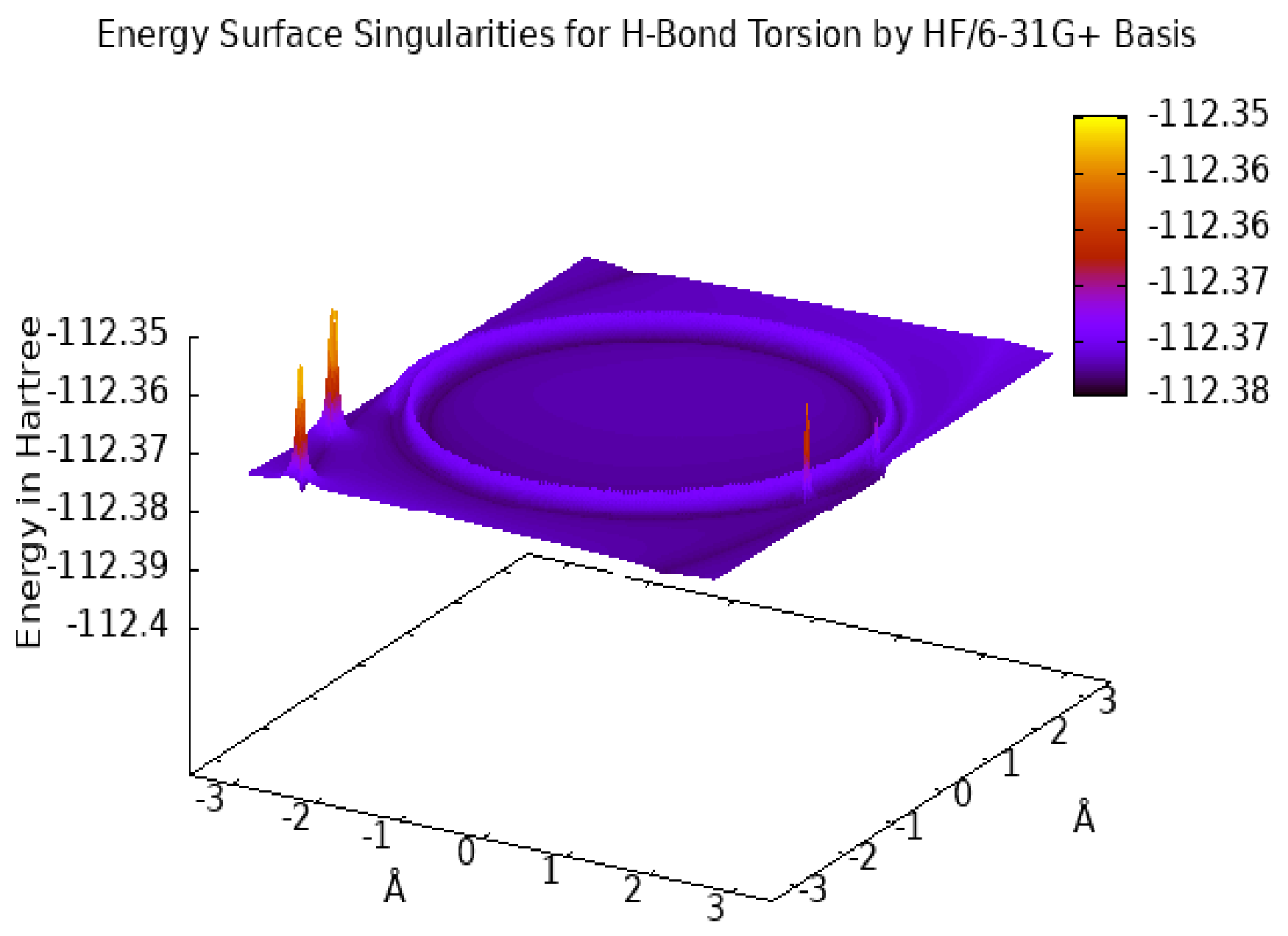

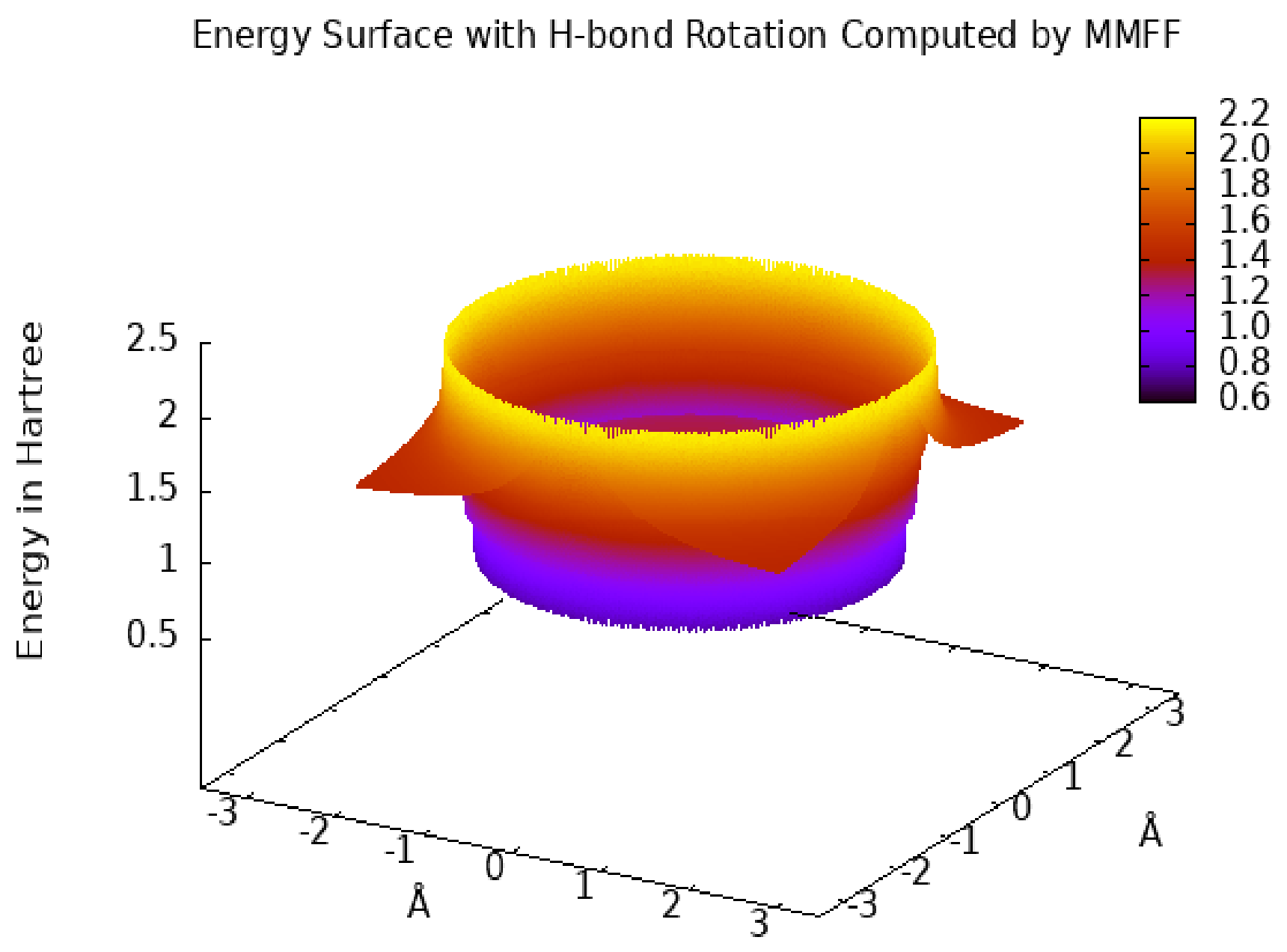